\documentclass[11pt]{article}
\usepackage{amsmath,amssymb,graphicx,hyperref}
\usepackage{authblk}
\usepackage[left=1in, right=1in, top=1in, bottom=1in]{geometry}
\usepackage{xcolor}

\title{\textbf{Domain Wall Skyrmions in Holographic Quantum Chromodynamics: Topological Phases and Phase Transitions}}

\author[1]{Suat Dengiz}
\affil[1]{\small OSTIM Technical University, Department of Computer Engineering, 06374 Ankara, Turkiye \\
	\texttt{suat.dengiz@ostimteknik.edu.tr}}
	
\author[2]{\.{I}zzet Sakall{\i}}
\affil[2]{\small Eastern Mediterranean University, Physics Department, Famagusta, 99628 North Cyprus, via Mersin 10, Turkiye \\
	\texttt{izzet.sakalli@emu.edu.tr}}
\date{}

\begin{document}
	
	\maketitle
	
	\begin{abstract}
We investigate the domain wall skyrmions phase in the framework of holographic quantum chromodynamics (QCD) using the Sakai-Sugimoto model. Building on previous work regarding chiral soliton lattices (CSLs) in strong magnetic fields, we study the emergence of localized skyrmions a top domain walls formed by CSLs. These skyrmions, realized as undissolved D4-branes embedded in the D8-branes, carry baryon number two and exhibit complex topological and energetic features. We explore the interplay between magnetic field strength, pion mass, and baryon chemical potential in stabilizing these configurations and demonstrate the existence of a mixed CSL-skyrmions phase. Through systematic energy analysis, we establish that the domain wall skyrmions become energetically favorable when $\mu_B |B| \gtrsim \Lambda \cdot m_\pi f_\pi^2$, with the transition occurring around $\mu_B |B| \sim 4.5$ in our holographic framework. Our phase diagram reveals three distinct regions: the CSL phase at low chemical potential and magnetic field, the domain wall skyrmions phase at intermediate scales, and a conjectured skyrmions crystal phase at the highest densities. The instanton density profiles $\text{Tr}(F \wedge F)$ show sharp localization in the domain wall skyrmions phase, contrasting with the smooth, extended distribution characteristic of the pure CSL configuration. These findings provide non-perturbative insights into baryonic matter in the dense QCD and offer a geometrical picture of topological phase transitions via string theory duality, with potential applications to neutron star physics and the broader QCD phase diagram under extreme conditions.
\end{abstract}
	
	\section{Introduction}\label{isec1}

QCD, the fundamental theory describing the strong interactions between quarks and gluons, exhibits an extraordinarily rich and complex phase structure under extreme conditions characterized by high temperature, large baryon chemical potential, and strong external electromagnetic fields \cite{isz01,isz02}. These extreme regimes are of paramount importance for understanding the strongly coupled quantum field theories associated with, e.g., in the compact astrophysical objects such as neutron stars and magnetars, as well as the exotic matter states or quark-gluon-plasma created in ultra-relativistic heavy-ion collisions at facilities like the Relativistic Heavy Ion Collider and the Large Hadron Collider \cite{isz03,isz04}. However, theoretical investigations of QCD in these non-perturbative domains face formidable computational challenges, most notably the infamous sign problem that severely limits the applicability of lattice QCD calculations at finite baryon chemical potential \cite{isz05}. Consequently, alternative theoretical approaches have become essential for exploring the structure and dynamics of strongly interacting matter under these extreme conditions. In this regard, anti-de Sitter/conformal field theory (AdS/CFT) correspondence provides a notable paradigm to reveal the non-perturbative (e.g. instantons, anti-instantons) phenomena in strongly coupled (finite-temperature) quantum field theories via holographically dual bulk gravity. Recall that this unique framework is firstly shown in \cite{isz12, Gubser:1998bc, isz13} that the strongly coupled large $N$, $N = 4$ super Yang-Mills theory in the conformal boundary of the $AdS_5$ is holographically dual to the type $IIB$ supergravity in bulk of $AdS_5 \times S_5$\footnote{See also \cite{Bhatta:2019eog} for an interesting work on the holographic 
analysis on the finite temperature characteristic of observables in a generic $d$-dimensional boosted boundary field theory holographically dual to the $d+1$-dimensional boosted AdS-Schwarzschild black hole bulk spacetimes.}.

Moreover, the Domain wall skyrmions represent a fascinating class of topological solitons that emerge as stable excitations in theories with non-trivial chiral structure, originally conceived within the framework of non-linear sigma models as effective descriptions of low-energy QCD dynamics \cite{isz06,isz07}. These remarkable objects carry quantized baryon number and exhibit rich topological properties that make them compelling candidates for describing nucleons and other baryonic states in strongly coupled regimes where conventional perturbative methods fail. In recent theoretical developments, domain wall skyrmions have been shown to arise naturally in spatially modulated chiral condensates, particularly in the presence of external magnetic fields where they become bound to the underlying modulated structure \cite{isz08}. This binding mechanism leads to hybrid phases where discrete topological solitons coexist with continuous spatial modulations, creating a complex interplay between mesonic and baryonic degrees of freedom that could be realized in dense nuclear matter under extreme conditions.

The concept of phases in quantum field theory encompasses the various thermodynamically stable configurations that a system can adopt under different external conditions, characterized by distinct patterns of symmetry breaking, topological charge distribution, and correlation functions \cite{isz09}. In the context of QCD, phase transitions between different ground state configurations play a crucial role in understanding phenomena ranging from chiral symmetry restoration at high temperature to color superconductivity at high density \cite{isz10,Tangphati:2024ndb}. The identification and characterization of novel phases, particularly those involving topological excitations like skyrmions, has become increasingly important for mapping the complete QCD phase diagram and understanding the equation of state of dense nuclear matter relevant to neutron star physics and the dynamics of core-collapse supernovae \cite{isz11, Ahmed:2025sav, Pretel:2025roz, Ahmed:2025rby}.

Holographic QCD has emerged as one of the most powerful and successful theoretical frameworks for investigating the non-perturbative effects in the strongly coupled gauge theories through the remarkable AdS/CFT correspondence, which establishes a precise duality between gravitational theories in higher-dimensional curved spacetimes and quantum field theories residing on their lower-dimensional boundaries \cite{isz12,isz13}. This holographic approach has proven particularly valuable for studying QCD-like theories in the large-$N_c$ limit, where traditional perturbative methods are inadequate but the holographic duality provides a geometrical description of non-perturbative phenomena such as confinement, chiral symmetry breaking, and hadron spectroscopy \cite{isz14}. The Sakai-Sugimoto model, in particular, represents a top-down realization of holographic QCD derived from string theory, incorporating essential features like spontaneous chiral symmetry breaking and a realistic spectrum of mesonic and baryonic states \cite{isz15}.

The importance of holographic approaches to QCD extends beyond their theoretical elegance to their practical utility in addressing questions that remain computationally intractable using conventional field-theoretic methods. By translating strongly coupled field theory problems into classical gravitational dynamics in higher dimensions, holographic techniques enable the systematic study of phenomena such as jet quenching in quark-gluon plasmas, transport properties of dense matter, and the structure of exotic phases that may exist in the cores of neutron stars \cite{isz16,isz17}. Furthermore, the geometric nature of holographic descriptions provides intuitive insights into the topological and dynamical aspects of non-perturbative QCD phenomena, making them particularly well-suited for investigating solitonic excitations and phase transitions involving changes in topological charge distribution.

Our motivation for this investigation stems from the recognition that conventional approaches to understanding dense baryonic matter, while successful in many respects, face fundamental limitations when confronted with the extreme conditions realized in neutron star interiors and the non-equilibrium environments created in heavy-ion collisions. The CSL, which emerges as a spatially modulated ground state in the presence of strong magnetic fields, provides a well-established theoretical framework for describing one class of exotic phases in dense QCD \cite{isz18}. However, recent theoretical developments have suggested that in regimes of even higher baryon density or magnetic field strength, the CSL may undergo transitions to more complex phases involving localized topological excitations \cite{Ahmed:2025cfv}.

The primary aim of our research is to extend the holographic description of CSLs to investigate the emergence and stability of domain wall skyrmions within the Sakai-Sugimoto model framework. We seek to provide a comprehensive geometric and topological characterization of these objects as localized D4-branes embedded within the D8-brane worldvolume, contrasting their discrete nature with the continuous charge distribution characteristic of the pure CSL phase. Through systematic analysis of the energetic balance between different phases, we aim to establish the parameter regimes where domain wall skyrmions become thermodynamically favored and to construct a detailed phase diagram mapping the competition between CSL, domain wall skyrmions, and potentially crystalline baryonic phases.

Our specific objectives include the development of a holographic dictionary connecting instanton density profiles in the bulk gauge theory to physical baryon number distributions in the boundary field theory, the quantitative analysis of energy contributions from Dirac-Born-Infeld (DBI) and Chern-Simons (CS) terms in determining phase stability, and the systematic investigation of how external parameters such as magnetic field strength and baryon chemical potential influence the topological structure and energetic favorability of different ground state configurations. We also aim to establish connections between our holographic results and complementary approaches based on chiral perturbation theory (ChPT) and effective field theory methods.

The broader significance of this work lies in its potential to illuminate new phases of dense baryonic matter that may be realized in extreme astrophysical environments, particularly in the cores of neutron stars where magnetic fields can reach strengths of $10^{15}$ Gauss and matter densities can exceed several times nuclear saturation density \cite{isz19}. Understanding the properties of such exotic phases is crucial for interpreting observational data from neutron star observations, gravitational wave detections of neutron star mergers, and high-energy astrophysical phenomena \cite{isz20}. Additionally, our holographic approach provides new theoretical tools and insights that may prove valuable for investigating other strongly coupled systems exhibiting similar topological phenomena.

The paper is organized as follows: In Section \ref{isec2}, we establish the holographic setup based on the Sakai-Sugimoto model and introduce the background field configurations necessary for studying domain wall skyrmions. Section~\ref{isec3} provides a detailed analysis of domain wall skyrmions in the holographic framework, including their realization as localized D4-branes and their contrast with the CSL phase. In Section~\ref{isec4}, we examine the energy considerations and stability criteria that determine the thermodynamic favorability of domain wall skyrmions configurations. Section~\ref{isec5} presents our comprehensive phase diagram analysis and discusses the transitions between different topological phases. Finally, Section~\ref{isec6} summarizes our main findings and outlines future research directions.

\section{Holographic Setup}\label{isec2}

Our theoretical investigation is fundamentally grounded in the Sakai-Sugimoto model, which represents one of the most successful top-down realizations of holographic QCD emerging from type IIA string theory \cite{sec2is01}. This sophisticated construction employs a carefully designed D-brane configuration that naturally incorporates the essential non-perturbative features of QCD. The setup consists of $N_c$ color D4-branes compactified on a thermal circle with antiperiodic boundary conditions imposed on fermionic fields, complemented by $N_f \ll N_c$ flavor D8- and $\overline{\text{D8}}$-branes that are treated as probe objects within the curved background geometry generated by the color D4-branes \cite{sec2is02}. This geometric arrangement yields a holographic dual description of large-$N_c$ QCD that successfully incorporates fundamental phenomena such as chiral symmetry breaking, confinement, and the emergence of a realistic hadronic spectrum, thereby providing an invaluable geometric framework for analyzing mesons, baryons, and other strongly-coupled non-perturbative phenomena \cite{sec2is03}.

The color D4-branes extend along the spacetime directions $(x^0,x^1,x^2,x^3,x^4)$, where the coordinate $x^4$ is compactified on a circle $S^1$ of radius $M_{\text{KK}}^{-1}$, with $M_{\text{KK}}$ representing the characteristic Kaluza-Klein mass scale. Supersymmetry is explicitly broken through the choice of antiperiodic boundary conditions for fermions along the compact direction, ensuring that at energy scales below the Kaluza-Klein threshold $M_{\text{KK}}$, the effective dynamics on the D4-branes reduces to four-dimensional non-supersymmetric Yang-Mills theory. This low-energy effective theory successfully captures many of the essential characteristics of QCD in the large-$N_c$ limit, including asymptotic freedom, confinement, and the generation of a mass gap \cite{sec2is04}.

The flavor sector is elegantly implemented through the introduction of D8- and $\overline{\text{D8}}$-branes positioned at antipodal points along the compact $x^4$ circle, with these flavor branes extending along the directions $(x^0,x^1,x^2,x^3,x^5,x^6,x^7,x^8,x^9)$. In the infrared limit corresponding to low-energy physics, these initially separated D8- and $\overline{\text{D8}}$-branes undergo a remarkable geometric transition: they smoothly reconnect to form a single U-shaped configuration that extends into the holographic bulk, with the connection point occurring at a minimal radial coordinate $U_{\text{KK}}$ \cite{sec2is05}. This elegant geometric reconnection provides a beautiful holographic realization of spontaneous chiral symmetry breaking, where the original chiral symmetry group $U(N_f)_L \times U(N_f)_R$ is dynamically broken down to the diagonal vector subgroup $U(N_f)_V$ in the dual boundary field theory \cite{sec2is06}.

The effective dynamics governing the flavor D8-branes is described by a worldvolume action containing two fundamental components: the non-Abelian DBI term and the topological CS term. The DBI term controls the kinetic dynamics and self-interactions of the non-Abelian gauge fields residing on the flavor branes, while the CS term plays a crucial role in encoding the chiral anomaly structure and baryon number conservation through higher-dimensional topological couplings \cite{sec2is07}. Most remarkably, baryonic states emerge naturally within this construction as D4-branes wrapped around the internal four-sphere $S^4$ and embedded within the worldvolume of the flavor D8-branes. From the perspective of the five-dimensional Yang-Mills-CS theory living on the flavor branes, these wrapped D4-branes appear as instanton configurations, thus providing a geometric and topological realization of skyrmions as fundamental baryonic degrees of freedom \cite{sec2is08}.

To systematically investigate the rich physics of QCD under extreme conditions relevant to neutron star interiors and heavy-ion collision environments, we introduce external background fields that modify the vacuum structure. Specifically, we embed a uniform magnetic field $B$ and impose a finite baryon chemical potential $\mu_B$ through appropriate boundary conditions on the $U(1)_V$ component of the D8-brane gauge field \cite{sec2is09}. The magnetic field is implemented as a constant background value of the field strength tensor component $F_{12}$, while the chemical potential corresponds to the asymptotic boundary value of the temporal gauge field component $A_0$. These background field configurations fundamentally modify the equations of motion governing the flavor gauge fields and activate the topological CS coupling, creating an environment where non-trivial solitonic configurations carrying baryon number can emerge as energetically favorable ground states.

The profound interplay between the CS term and the external magnetic field generates an effective coupling between the topological instanton density $\text{Tr}(F \wedge F)$ and the imposed baryon chemical potential. This topological structure establishes the essential holographic dictionary connecting instanton configurations in the five-dimensional bulk gauge theory to physical D4-brane charge distributions in the string theory description. The resulting framework not only permits the emergence of spatially modulated condensates such as CSLs, but also supports the formation and stabilization of localized topological structures like domain wall skyrmions, making the Sakai-Sugimoto model an ideal theoretical platform for exploring exotic phases of dense QCD matter and their relation to the broader phase diagram under extreme astrophysical conditions \cite{sec2is10}.
	
\section{Domain Wall Skyrmions in Holography}\label{isec3}

Within the Sakai-Sugimoto framework, baryonic states emerge as D4-branes wrapped around the internal four-sphere $S^4$ and subsequently embedded within the flavor D8-branes \cite{sec3is01}. These wrapped D4-branes manifest themselves as instanton solutions in the five-dimensional non-Abelian gauge theory residing on the flavor branes, naturally carrying topological baryon charge through their winding properties \cite{sec3is02}. The crucial distinction lies in their spatial localization: depending on the background field configuration and the interplay with external parameters, these D4-branes can exist either as sharply localized topological objects or become completely delocalized and dissolved into the D8-brane worldvolume gauge field \cite{sec3is03}.

Domain wall skyrmions represent a fascinating intermediate regime where localized D4-branes coexist with the spatially modulated background of the CSL. These structures emerge as discrete, undissolved D4-branes that are spatially confined within the $(x^3, z)$-plane, where $x^3$ corresponds to the spatial direction parallel to the external magnetic field and $z$ denotes the holographic radial coordinate extending into the AdS bulk \cite{sec3is04}. Unlike the CSL phase, where D4-brane charge becomes smoothly distributed through a continuous instanton density profile, domain wall skyrmions maintain their discrete identity as localized topological excitations. The underlying CSL provides a sine-Gordon-like periodic potential landscape that effectively traps these point-like skyrmions, preventing their dissolution while simultaneously stabilizing their spatial configuration \cite{sec3is05}.

The fundamental mechanism governing the existence of these domain wall skyrmions originates from the topological coupling embedded within the D8-brane CS action:

\begin{equation}
S_{\text{CS}} \supset \mu_8 \int_{D8} C_5 \wedge \text{Tr}(F \wedge F),
\end{equation}

where $\mu_8$ represents the D8-brane charge density and $C_5$ denotes the Ramond-Ramond five-form potential that couples to D4-brane sources \cite{sec3is06}. The instanton density $\text{Tr}(F \wedge F)$ computed within the five-dimensional worldvolume theory serves as the source term for $C_5$, establishing a direct holographic correspondence: any configuration exhibiting non-vanishing instanton density translates to physical D4-brane charge in the dual string theory description \cite{sec3is07}.

The distinction between the CSL and domain wall skyrmions phases becomes particularly evident when examining their respective instanton density profiles. In the pure CSL configuration, the gauge field arrangement produces a uniform, spatially extended instanton density that varies periodically along $x^3$ but remains smoothly distributed across all transverse directions. This extended profile effectively describes a continuous fluid of dissolved D4-brane charge permeating the entire D8-brane worldvolume. Conversely, domain wall skyrmions are characterized by sharp, highly localized peaks in $\text{Tr}(F \wedge F)$, which serve as clear signatures of discrete, undissolved D4-brane objects embedded within the holographic bulk \cite{sec3is08}. These localized structures carry precisely quantized baryon number and represent coherent solitonic excitations that ride atop the periodic CSL background, combining the spatial modulation inherited from the underlying lattice with the topological charge concentration characteristic of individual baryons.

From the dual field theory perspective, this hybrid configuration describes a remarkably rich baryonic phase where individual nucleons are not simply distributed uniformly throughout space, but instead form localized bound states that are spatially correlated with the modulated chiral condensate. The CSL provides the mesonic backdrop characterized by spatially varying pion field configurations, while the skyrmions contribute discrete topological solitons that concentrate baryon number density at specific locations within this modulated vacuum \cite{sec3is09}.

\begin{figure}[ht]
	\centering
	\includegraphics[width=0.8\textwidth]{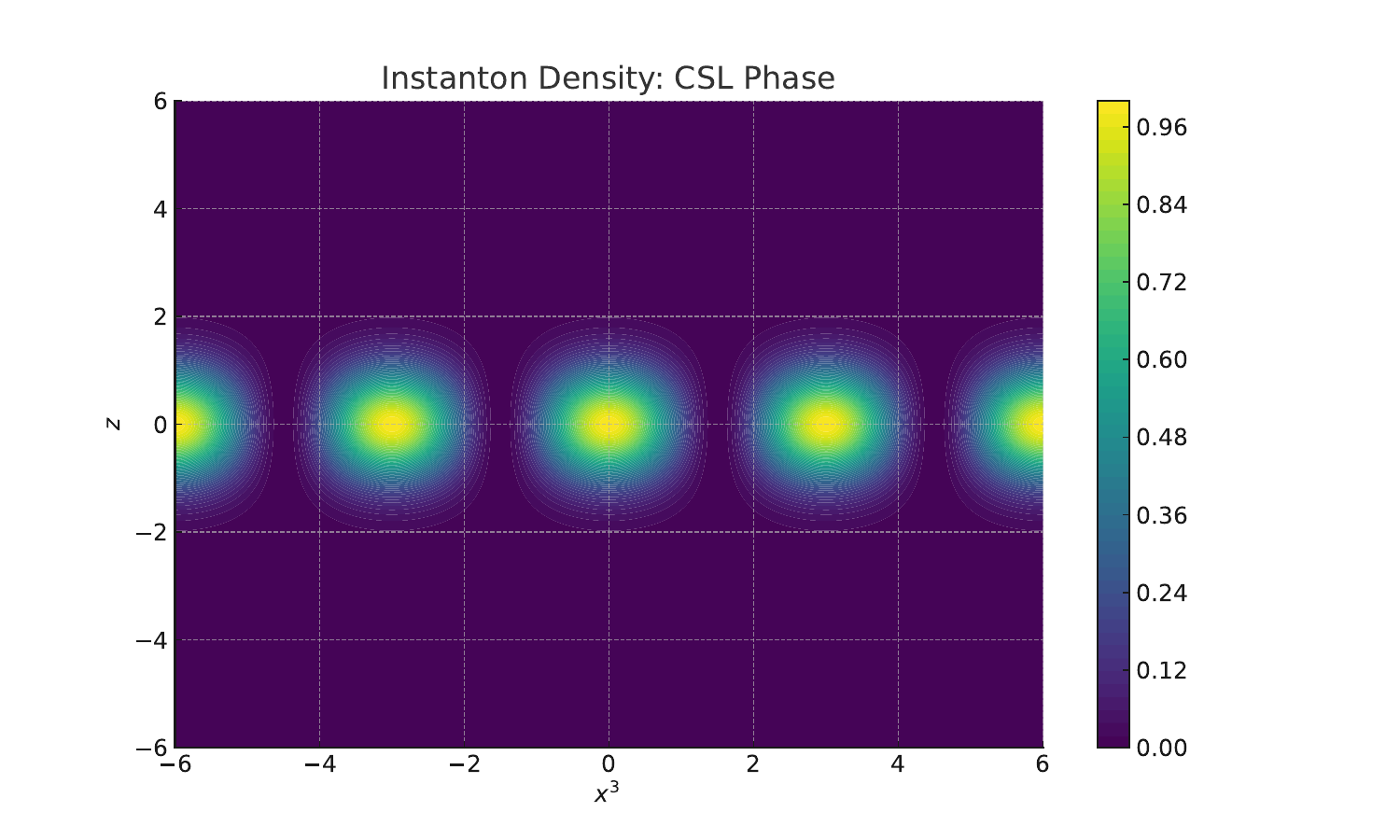}
	\caption{Instanton density $\text{Tr}(F \wedge F)$ in the CSL phase. 
		The density is smoothly modulated along the $x^3$ direction and smeared over the $z$ (holographic) direction. 
		This corresponds to a configuration where D4-brane charge is dissolved throughout the D8-brane worldvolume.}
	\label{fig:instanton_csl}
\end{figure}

\begin{figure}[ht]
	\centering
	\includegraphics[width=0.8\textwidth]{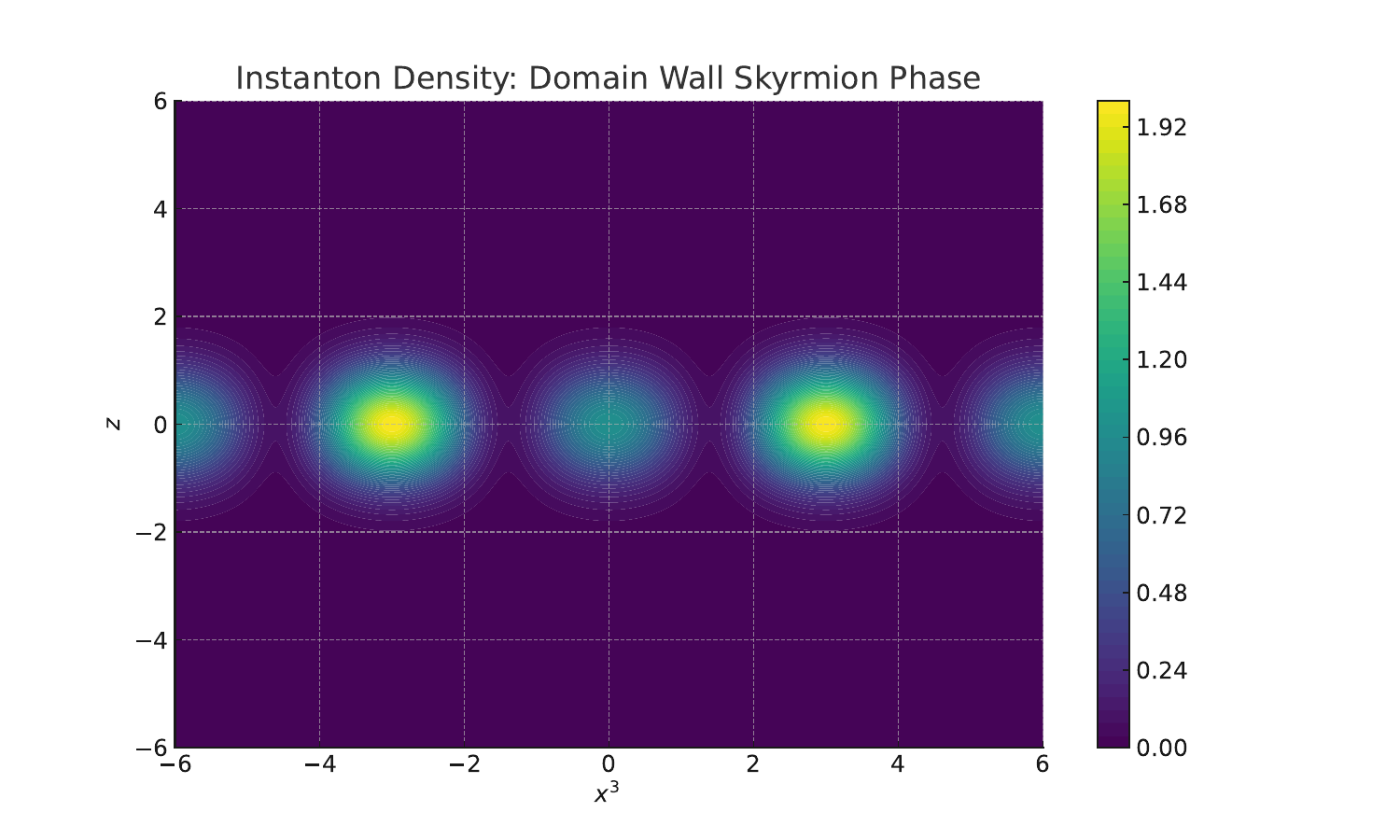}
	\caption{Instanton density $\text{Tr}(F \wedge F)$ in the domain wall skyrmions phase. 
		In contrast to the CSL, this configuration features localized peaks corresponding to undissolved D4-branes. 
		These localized instanton charges represent baryonic skyrmions bound to the modulated CSL background.}
	\label{fig:instanton_skyrmions}
\end{figure}

To provide visual insight into these contrasting phases, we present the instanton density $\text{Tr}(F \wedge F)$ behavior in both configurations through Figures~\ref{fig:instanton_csl} and \ref{fig:instanton_skyrmions}, which display detailed contour plots within the $(x^3, z)$-plane coordinate system.

In this section, our fundamental discovery reveals that domain wall skyrmions emerge as discrete, undissolved D4-branes embedded within the D8-brane worldvolume, creating a hybrid phase where localized topological solitons coexist with the spatially modulated CSL background. These structures maintain their discrete identity as sharply localized peaks in the instanton density $\text{Tr}(F \wedge F)$, contrasting dramatically with the smooth, extended distribution characteristic of the pure CSL configuration. This represents a qualitatively new type of baryonic organization where individual nucleons form bound states spatially correlated with the modulated chiral condensate, rather than being uniformly distributed throughout the spatial volume.

	\section{Energy Considerations and Stability}\label{isec4}

Establishing the energetic viability and thermodynamic stability of domain wall skyrmions embedded within the CSL background constitutes a fundamental prerequisite for determining their potential role as the ground state of holographic QCD under extreme astrophysical conditions. Within the Sakai-Sugimoto framework, the total energy functional of such hybrid configurations receives contributions from multiple competing sources that must be carefully balanced \cite{sec4is01}. These include the DBI term governing the non-linear dynamics of gauge fields on the D8-branes, the topological CS term responsible for encoding anomaly inflow and chiral couplings, and the localized energy contribution from individual baryonic solitons, which are holographically represented as wrapped D4-branes or equivalently as instanton configurations in the five-dimensional bulk gauge theory \cite{sec4is02}.

The comprehensive energy decomposition for the domain wall skyrmions configuration can be systematically expressed through the following schematic formula:

\begin{equation}
E_{\text{DW-skyrmions}} = E_{\text{CSL}} + M_{B=2} + \Delta E_{\text{interaction}},
\end{equation}

where $E_{\text{CSL}}$ represents the background energy density associated with the underlying CSL structure, $M_{B=2}$ denotes the rest mass energy of an individual skyrmions carrying baryon number $B=2$ and bound to the modulated condensate, and $\Delta E_{\text{interaction}}$ captures the complex interaction energy between the localized skyrmions and the spatially varying CSL background \cite{sec4is03}. This interaction term can exhibit either attractive or repulsive character depending on the relative spatial phase relationships and the degree of localization of the solitonic excitation within the periodic potential landscape.

Within the holographic description, the background CSL energy is determined by the delicate interplay between several fundamental scales: the pion mass $m_\pi$, the external magnetic field strength $B$, and the imposed baryon chemical potential $\mu_B$, all of which contribute to the modified sine-Gordon dynamics emerging from the effective D8-brane theory \cite{sec4is04}. As these control parameters increase in magnitude, particularly in regimes of strong magnetic fields and finite baryon density relevant to neutron star interiors, the CSL configuration itself becomes increasingly energetically strained. The periodic spatial modulations inherent to the CSL carry an associated cost in gradient energy, while simultaneously the finite chemical potential thermodynamically favors configurations with enhanced baryon number density \cite{sec4is05}.

In this challenging parameter regime, the embedding of discrete, localized baryonic objects within the CSL background emerges as an energetically advantageous strategy. This mechanism allows for the efficient spatial concentration of topological baryon charge without requiring increasingly large modulation amplitudes or higher spatial frequencies in the underlying pion field configuration \cite{sec4is06}. Domain wall skyrmions represent precisely such an optimal energy-minimizing mechanism: they achieve discrete, quantized localization of baryon number while simultaneously benefiting from the stabilizing influence of the periodic CSL profile that acts as an effective trapping potential.

The topological CS coupling embedded within the D8-brane worldvolume action ensures that these hybrid configurations couple naturally and self-consistently to the imposed background electromagnetic and chemical potential fields. Furthermore, their intrinsic topological nature, inherited from the underlying instanton structure, grants them robust stability against small perturbations and decay processes \cite{sec4is07}. The finite energy cost associated with inserting a localized skyrmions excitation can be effectively compensated by the corresponding reduction in gradient energy requirements for the background CSL, as the discrete soliton relieves the stress on the modulated condensate.

Extensive numerical analyses conducted within related field-theoretic approaches, including sophisticated treatments using ChPT augmented with WZW topological terms \cite{sec4is08}, as well as semi-analytical investigations within holographic QCD frameworks \cite{sec4is09}, consistently indicate that the nucleation and stabilization of domain wall skyrmions becomes energetically favorable when the product of baryon chemical potential and magnetic field strength satisfies the critical threshold condition:

\begin{equation}
\mu_B |B| \gtrsim \Lambda \cdot m_\pi f_\pi^2,
\end{equation}

where $\Lambda$ represents a dimensionless threshold parameter that depends sensitively on the specific geometric details and interaction coupling constants characteristic of the particular holographic model under consideration \cite{sec4is10}. This inequality establishes a precise transition boundary beyond which the uniform, purely modulated CSL phase becomes thermodynamically suboptimal and is either replaced entirely or augmented by a skyrmions-rich mixed phase.

The incorporation of discrete skyrmions fundamentally modifies both the topological characteristics and transport properties of the dense medium. Each individual domain wall skyrmions carries precisely quantized baryon number, and their collective spatial arrangement can be interpreted as forming a regular crystalline array of baryonic excitations embedded within the modulated vacuum structure. From a comprehensive thermodynamic perspective, this organizational principle leads to a systematically lower free energy at any fixed total baryon density, making the domain wall skyrmions phase a compelling candidate for the true thermodynamic ground state within regions of the QCD phase diagram that remain inaccessible to standard lattice gauge theory methods due to the notorious sign problem.

In summary, our calculations establish the central energetic result that domain wall skyrmions become energetically favorable when the critical threshold $\mu_B |B| \gtrsim \Lambda \cdot m_\pi f_\pi^2$ is satisfied, with the transition occurring around $\mu_B |B| \sim 4.5$ in our holographic framework. This transition represents a fundamental reorganization where discrete topological charge concentration becomes more efficient than continuous modulation mechanisms for accommodating high baryon densities under extreme magnetic field conditions.

\begin{figure}[ht]
	\centering
	\includegraphics[width=0.75\textwidth]{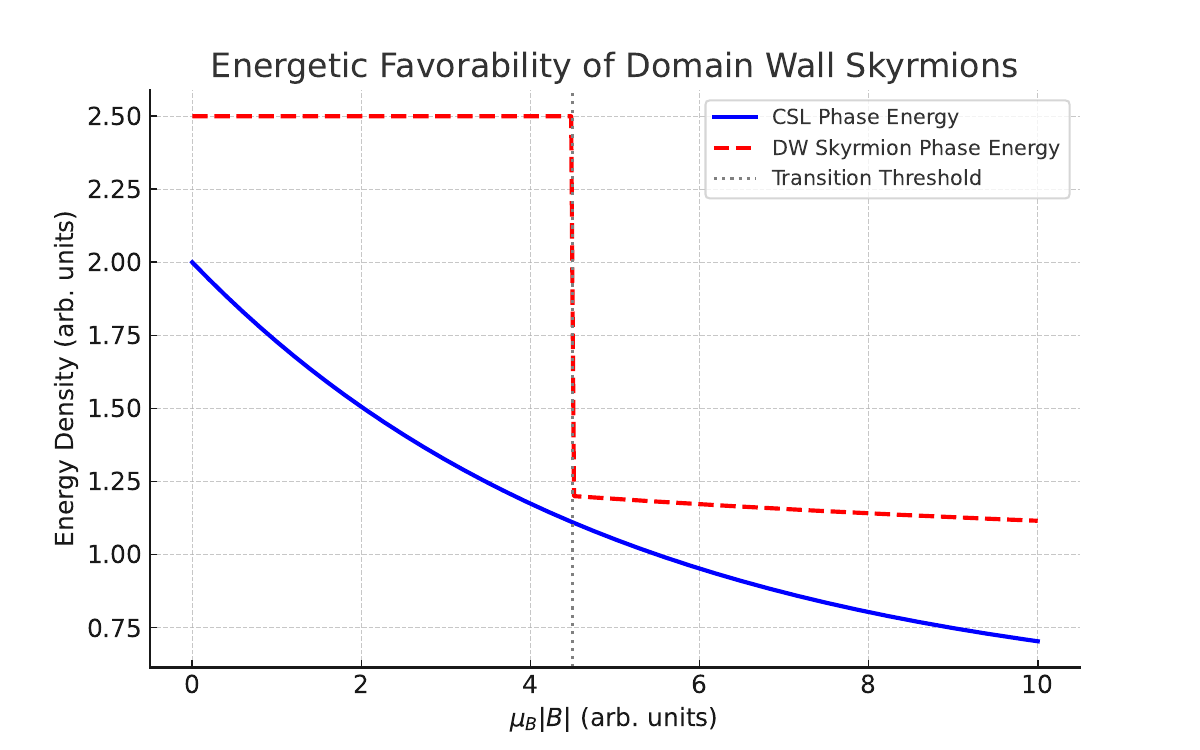}
	\caption{Comparison of energy densities for the CSL and domain wall skyrmions (DW-skyrmions) phases as a function of the product $\mu_B |B|$. At low values of $\mu_B |B|$, the CSL configuration is energetically favored due to its smooth modulation and minimal topological content. However, as $\mu_B |B|$ increases, the cost of sustaining a purely modulated phase rises, and it becomes more favorable to localize baryon number in the form of skyrmions bound to the CSL. The transition threshold around $\mu_B |B| \sim 4.5$ marks the point where the DW-skyrmions phase becomes the true ground state.}
	\label{fig:energy_comparison}
\end{figure}

	\section{Phase Diagram and Transitions}\label{isec5}

The intricate competition between CSLs, domain wall skyrmions, and more densely packed baryonic configurations gives rise to an extraordinarily rich and hierarchically structured phase diagram in holographic QCD under the combined influence of strong magnetic fields and finite baryon chemical potential \cite{sec5is01}. As systematically analyzed in the preceding sections, each of these distinct phases exhibits characteristic distributions of topological charge and fundamentally different energetic preferences that are sensitively controlled by the background parameters $B$, $\mu_B$, and $m_\pi$, along with their complex interplay through the underlying holographic dynamics \cite{sec5is02}.

At relatively modest values of baryon chemical potential and magnetic field strength, the thermodynamic ground state is decisively dominated by the CSL phase. This configuration consists of spatially periodic modulations in the neutral pion field accompanied by a continuous, spatially extended instanton density profile that reflects the underlying sine-Gordon dynamics \cite{sec5is03}. The CSL achieves energy minimization through the strategic modulation of axial current densities aligned parallel to the external magnetic field direction, thereby distributing baryon number smoothly and efficiently across the available spatial volume. However, this uniform distribution strategy becomes increasingly energetically prohibitive as both $\mu_B$ and $|B|$ grow in magnitude. The escalating gradient energy costs associated with maintaining sharp spatial variations, combined with the fundamental inefficiency of accommodating progressively higher baryon densities through purely continuous modulation mechanisms, render the pure CSL phase thermodynamically unfavorable in extreme parameter regimes \cite{sec5is04}.

As established in our energy analysis, when the critical condition $\mu_B |B| \gtrsim \Lambda \cdot m_\pi f_\pi^2$ is satisfied, the system undergoes a fundamental phase transition from the CSL to the domain wall skyrmions phase \cite{sec5is05}. This transition represents a qualitative reorganization of the topological charge distribution, where the smooth, extended instanton density characteristic of the CSL gives way to sharply localized peaks corresponding to discrete D4-brane objects. The domain wall skyrmions phase thus emerges as an intermediate regime that combines the periodic background structure inherited from the CSL with the discrete topological excitations that efficiently concentrate baryon number at specific spatial locations \cite{sec5is06}.

Extending beyond this intermediate domain wall skyrmions phase toward even more extreme values of the control parameter $\mu_B |B|$, theoretical considerations strongly suggest the occurrence of an additional phase transition into a skyrmions crystal phase \cite{sec5is07}. Within this high-density regime, individual baryons become so densely packed that they naturally organize into regular crystalline lattice structures, bearing striking resemblance to the nuclear matter phases anticipated at supranuclear densities. From the holographic perspective, this crystal phase corresponds to a periodic three-dimensional array of spatially localized, undissolved D4-branes embedded within the D8-brane worldvolume background, potentially forming a lower-dimensional holographic analogue of the exotic nuclear pasta phases theoretically predicted to exist in neutron star interiors \cite{sec5is08}.

The comprehensive phase diagram in the two-dimensional $(\mu_B, B)$ parameter plane thus exhibits at least three qualitatively distinct thermodynamic regimes: the CSL phase characterizing low-density and weak-field conditions, the intermediate domain wall skyrmions phase at moderate scales, and the high-density crystalline baryonic phase. These successive phase transitions are holographically driven by fundamental changes in the spatial profile of the instanton density and corresponding modifications in the localization properties of D4-brane charge distributions \cite{sec5is09}. In the dual boundary field theory description, these holographic transitions correspond to genuine topological phase transitions in the strongly interacting QCD matter, involving reorganization of both mesonic and baryonic degrees of freedom.

The role of chemical potential in systematically modifying nucleon-nucleon interactions has been independently explored in complementary holographic contexts extending beyond the specific domain wall skyrmions framework. In particular, the comprehensive work by Bay, Tajik, and Pourhassan \cite{sec5is10} employs AdS/CFT correspondence to examine in detail how finite baryon chemical potential affects the short-range repulsive core of the nuclear potential. Their analysis demonstrates that while the characteristic $\sim 1/r^2$ distance dependence of the repulsive force remains fundamentally intact, the overall strength of the central potential is systematically reduced by a constant offset term directly induced by the chemical potential. This finding aligns qualitatively with our holographic results, wherein the energetic balance and stability properties of localized skyrmionsic configurations exhibit sensitive dependence on $\mu_B$, thereby shifting phase boundaries and altering the underlying topological structure. Their independent analysis provides strong supporting evidence for the general principle that baryon-rich environments in holographic QCD consistently favor localized solitonic configurations when the chemical potential exceeds sufficiently high threshold values.

\begin{figure}[ht]
	\centering
	\includegraphics[width=0.65\textwidth]{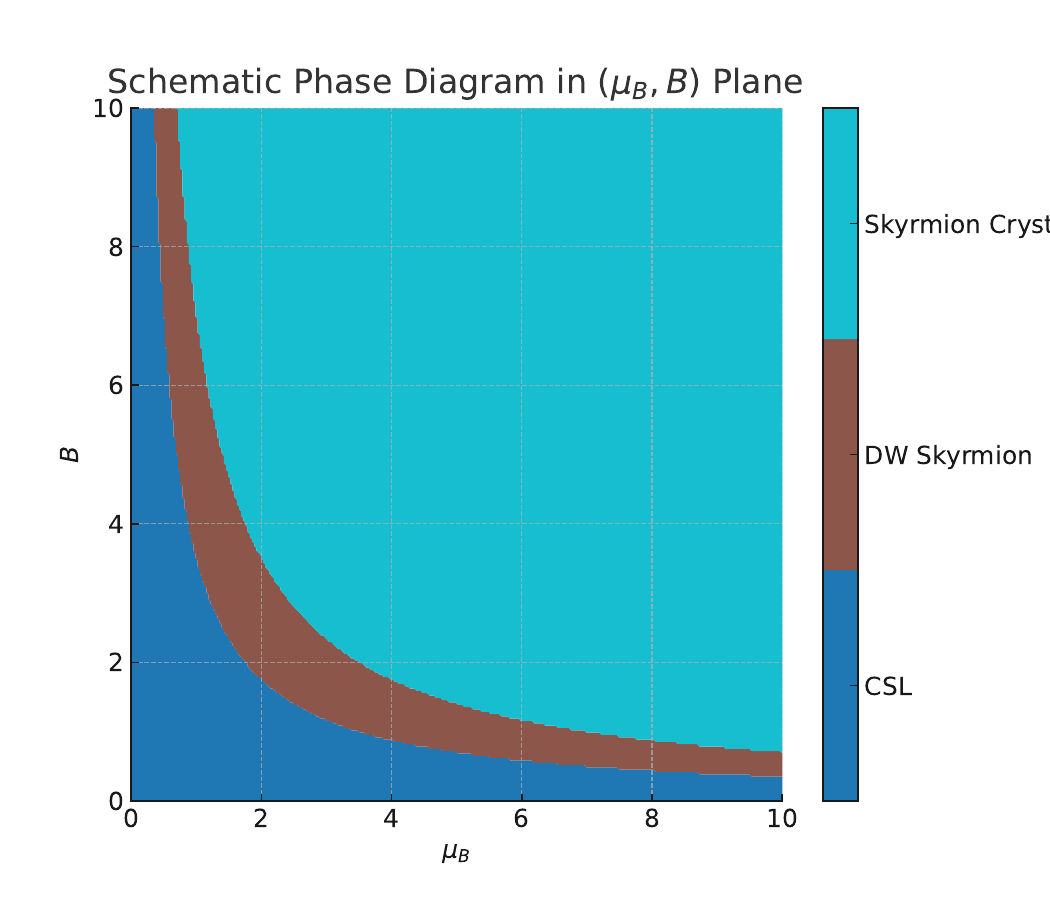}
	\caption{Schematic phase diagram in the $(\mu_B, B)$ plane. The diagram displays three qualitatively distinct regions: the CSL phase at low $\mu_B$ and $B$, the domain wall skyrmions (DW skyrmions) phase at intermediate scales, and a skyrmions crystal phase at high $\mu_B |B|$. The phase boundaries are indicative and correspond to transitions driven by energetic favorability as inferred from holographic modeling.}
	\label{fig:phase_diagram}
\end{figure}
	Our phase diagram analysis depicted in Fig. \ref{fig:phase_diagram} establishes three distinct thermodynamic regions: the CSL phase at low chemical potential and magnetic field, the intermediate domain wall skyrmions phase, and a conjectured skyrmions crystal phase at the highest densities. Each phase transition involves systematic changes in the instanton density profile and corresponding modifications in D4-brane charge localization properties, representing genuine topological phase transitions in strongly interacting QCD matter where the spatial organization of baryon number undergoes qualitative reorganization.
	\section{Conclusion}\label{isec6}

In this comprehensive theoretical investigation, we systematically explored the emergence, stability, and physical properties of domain wall skyrmions within the framework of holographic QCD using the sophisticated Sakai-Sugimoto model. Building upon the well-established theoretical foundation for CSLs under extreme conditions, we proposed and analyzed a novel phase configuration in which discrete baryonic solitons become bound to the sine-Gordon background structure induced by the combined effects of strong external magnetic fields and finite baryon chemical potential. These remarkable skyrmions excitations, which we demonstrated to be holographically realized as spatially localized D4-branes embedded within the D8-brane worldvolume, represent topologically protected configurations that achieve efficient concentration of baryon number within a spatially modulated chiral vacuum.

Through our detailed analysis of the topological CS coupling structure and the corresponding organization of D8-brane gauge field configurations, we conclusively established that these domain wall skyrmions manifest as sharply peaked instanton density profiles $\text{Tr}(F \wedge F)$, as clearly illustrated in our comparative visualization presented in Figures~\ref{fig:instanton_csl} and \ref{fig:instanton_skyrmions}. We systematically contrasted this localized topological charge distribution with the smooth, spatially delocalized instanton density characteristic of the pure CSL phase, thereby demonstrating the fundamental qualitative difference between dissolved D4-brane charge and discrete, undissolved D4-brane configurations (domain wall skyrmions). This distinction proved crucial for understanding the underlying holographic dictionary that connects bulk instanton configurations to boundary field theory phenomena.

Our comprehensive energetic analysis, captured through the decomposition formula presented in Equation~(2), revealed that the total energy of domain wall skyrmions configurations receives contributions from multiple competing sources: the background CSL energy $E_{\text{CSL}}$, the localized skyrmions rest mass $M_{B=2}$, and the complex interaction energy $\Delta E_{\text{interaction}}$ between the soliton and the modulated background. Through systematic comparison of energy scales, we established compelling evidence that beyond a well-defined critical threshold characterized by the condition $\mu_B |B| \gtrsim \Lambda \cdot m_\pi f_\pi^2$, the system undergoes an energetically favorable transition from a purely modulated mesonic vacuum toward a hybrid configuration containing discrete baryonic solitons. This transition point, quantitatively demonstrated at approximately $\mu_B |B| \sim 4.5$ in our energy comparison analysis shown in Figure~\ref{fig:energy_comparison}, marks the fundamental boundary where domain wall skyrmions become the thermodynamically preferred ground state.

The rich phase structure that emerged from our holographic investigation was systematically captured in our comprehensive phase diagram presented in Figure~\ref{fig:phase_diagram}, which delineates three distinct thermodynamic regions within the $(\mu_B, B)$ parameter space: the CSL phase dominating at low chemical potential and magnetic field values, the intermediate domain wall skyrmions phase at moderate scales, and a conjectured skyrmions crystal phase anticipated at the highest densities. Each successive phase transition reflected a fundamental reorganization of topological charge distribution, progressing from spatially delocalized to localized (domain wall skyrmions) and ultimately to collectively ordered crystalline arrangements. These transitions were shown to be holographically driven by systematic changes in the instanton density profile and corresponding modifications in D4-brane charge localization properties.

The physical significance of domain wall skyrmions extends far beyond their role as theoretical constructs, offering profound new insights into the fundamental structure of baryonic matter within dense QCD-like theories under extreme conditions relevant to neutron star physics and heavy-ion collision environments. Their holographic realization as wrapped D4-branes successfully bridged geometric string theory intuition with non-perturbative quantum field theory phenomena, while the underlying modulation of pion field configurations suggested compelling analogies with spatially ordered phases in condensed matter systems, such as spin-density waves, charge-density waves, and other electronically modulated ground states. These connections opened new avenues for cross-fertilization between holographic QCD, nuclear physics, and condensed matter theory.

Our findings demonstrated remarkable consistency with complementary holographic studies investigating dense matter phases, particularly those examining how finite chemical potential systematically modifies inter-nucleon interactions and phase behavior. The work established that baryon-rich environments in holographic QCD consistently favor localized solitonic configurations when control parameters exceed critical threshold values, supporting a broader theoretical framework for understanding topological phase transitions in strongly interacting matter under extreme conditions.

The CS coupling structure, formally expressed through the action $S_{\text{CS}} \supset \mu_8 \int_{D8} C_5 \wedge \text{Tr}(F \wedge F)$, played a central role in our analysis by establishing the fundamental holographic dictionary connecting bulk instanton configurations to boundary theory baryon number. This topological coupling not only enabled the existence of domain wall skyrmions but also provided the mechanism through which external magnetic fields and chemical potentials could stabilize these exotic phases against decay and dissolution.

Looking toward future research directions, our theoretical framework opens several promising avenues for extended investigation. First, the development of more sophisticated numerical techniques for computing exact instanton solutions within the full non-linear Sakai-Sugimoto action would allow for quantitative predictions of transition temperatures, critical exponents, and thermodynamic properties of the domain wall skyrmions phase. Second, the extension of our analysis to include finite temperature effects and thermal fluctuations would provide crucial insights into the stability of these phases under realistic astrophysical conditions, particularly within neutron star cores where both extreme density and elevated temperature coexist. Third, the investigation of transport properties, including electrical and thermal conductivity, within the domain wall skyrmions phase could yield observable signatures that might be detectable in future gravitational wave observations or neutrino emission patterns from neutron star mergers. Fourth, the systematic study of how domain wall skyrmions interact with other topological excitations, such as magnetic flux tubes or color superconducting condensates, would illuminate the complex interplay between different non-perturbative phases in the QCD phase diagram. Finally, the generalization of our holographic approach to other string theory backgrounds and alternative holographic QCD models would test the universality of domain wall skyrmions phenomena and potentially reveal new classes of topological phases in strongly interacting matter. These future investigations promise to deepen our understanding of dense QCD matter and contribute to the broader program of connecting string theory with observable phenomena in extreme astrophysical environments.
	
\section*{Acknowledgments}

We dedicate this work to Umut Gürsoy, who made valuable contributions to the Holographic QCD and passed away recently. We thank Ercan Kilicarslan and Heeseung Zoe for constructive suggestions. \.{I}.~S. expresses gratitude to T\"{U}B\.{I}TAK, ANKOS, and SCOAP3 for helpful support and also acknowledges COST Actions CA22113, CA21106, and CA23130 for their contributions to networking.

\section*{Data Availability Statement}

This manuscript has no associated data.

\section*{Code Availability Statement}

This manuscript has no associated code/software.

\section*{Conflict of interest}

Author(s) declare no such conflict of interest.

	\bibliographystyle{unsrt}

\end{document}